\documentclass[%
 reprint,
nofootinbib,
nobibnotes,
 amsmath,amssymb,
 aps,
]{revtex4-2}

\usepackage{graphicx}
\usepackage{dcolumn}
\usepackage{bm}

\newcommand{\md}{\mathrm{d}}

\newcommand{\me}{\mathrm{e}}

\newcommand{\nn}{\nonumber}

\newcommand{\pattern}{\Omega_{\mathrm{p}}}

\newcommand{\inlineratio}{rms $\delta J_R\,/\,$rms $\delta J_\varphi\,$}

\begin{document}

\preprint{APS/123-QED}

\title{Why is the Galactic disk so cool?}

\author{Chris Hamilton}
 \email{chamilton@ias.edu}
\affiliation{Institute for Advanced Study, Einstein Drive, Princeton NJ 08540
}%
\author{Shaunak Modak}
\affiliation{Department of Astrophysical Sciences, Princeton University, 4 Ivy Lane, Princeton NJ 08544
}%
\author{Scott Tremaine}
\affiliation{Institute for Advanced Study, Einstein Drive,, Princeton NJ 08540
}%

\date{\today}

\begin{abstract}
The bulk of old stars in the Galactic disk have migrated radially by up to several kpc in their lifetimes, yet the disk has remained relatively cool, i.e., the \textit{ratio} of radial heating to migration has been small. Here, we demonstrate that this small ratio places very strong constraints on which mechanisms could have been responsible for orbital transport in our Galaxy. For instance, Sellwood \& Binney's mechanism of nonlinear horseshoe transport by spirals tends to produce too high a ratio of heating to migration, unless the spirals' amplitudes are heavily suppressed away from their corotation resonances, or their pitch angles are significantly larger than is observed. This problem is only made worse if one includes the effect of the Galactic bar, diffusion due to disk or halo substructure, etc. Resonant (but non-horseshoe) scattering by spirals can drive transport consistent with the data, but even this requires some fine-tuning. In short, reproducing both the observed radial migration \textit{and} the small ratio of heating to migration is a highly nontrivial requirement, and poses a significant challenge to models of the Milky Way's dynamical history, theories of spiral structure, and the identification of `Milky Way analogues' in cosmological simulations.
\end{abstract}


\maketitle

\section{Introduction}
\label{sec:Introduction}

A central goal of Galactic astronomy in the era of surveys such as GAIA, APOGEE, etc., is to infer the assembly history and evolution of our Galaxy from the imprints left in the kinematics and chemistry of disk stars by certain `perturbers.' For example, features of the Hercules moving group have been interpreted as evidence for a slowing Galactic bar \citep{chiba2021resonance}, the vertical phase space snail has been tied to perturbations from infalling dwarf galaxies and/or dark matter subhaloes \citep{antoja2018dynamically, laporte2019footprints, tremaine2023origin}, and the Galactic disk's warp and flare have been linked both to satellite impacts and to possible asphericity or precession of the dark halo \citep{ostrikerbinney1989, poggio2018galactic, han2023tilted}.

However, a significant limitation of all such studies has been that \textit{we have little knowledge of the statistics of stochastic dynamical perturbations in the Galactic disk over the last several Gyr}. In other words, we have scant idea of the statistical properties of the spiral arms, molecular clouds, etc.\ that ought to accompany the `perturbers' referred to above in any realistic dynamical model. These stochastic fluctuations are important partly because they lead to direct orbital transport \citep{Binney1988-zy}, and partly because they can destroy delicate resonant effects upon which some of the aforementioned results rely \citep{hamilton2023galactic}. In lieu of an understanding of these fluctuations, modelers have had to resort to (i) ignoring them altogether (e.g., \cite{chiba2021resonance}), (ii) applying toy models with additional free parameters (e.g., \cite{tremaine2023origin}), or (iii) running full $N$-body simulations (e.g., \cite{laporte2019footprints}). None of these approaches is perfectly satisfying, because one can never be entirely confident that the signal in the data is not due, at least in part, to perturbations absent from the model.

Here, we take a major step toward constraining the aforementioned fluctuation spectrum by honing in on a single observational fact: namely, that the Galactic disk is \textit{cool}. More precisely, we focus on the fact that the old stars in the Milky Way's disk have been `churned' efficiently in orbital guiding radius $R_\mathrm{g}$ throughout their lifetimes --- often by several kpc in total --- while remaining on orbits of remarkably low eccentricity \citep{Frankel2020-vy}. It is even possible that the entire outer disk beyond $\sim 15$ kpc formed via radial migration of stars from smaller radii \citep{lian2022quantifying}. Said in more technical language, in the Milky Way's disk, radial migration (shuffling of angular momenta, a.k.a. azimuthal actions $J_\varphi$) is a \textit{significantly} more efficient process than radial heating (increasing the radial actions $J_R$).

To measure this tendency quantitatively, Frankel et al. \cite{Frankel2020-vy} constructed a parameterized model of radial migration and heating, assuming simple functional forms for the stellar distribution function (DF), and assuming that its evolution followed a Fokker-Planck equation.
Their model that best fit the GAIA/APOGEE data revealed that over the last $6$\,Gyr, Milky Way disk stars have roughly random-walked in guiding radius $R_\mathrm{g}$ according to rms\,\,$\delta R_\mathrm{g}(t) \sim 1\,\mathrm{kpc} \times \left(t/ 1 \, \mathrm{Gyr} \right)^{1/2}$. (Here, and throughout this paper, we define the rms change in a quantity $x$ as rms\,\,$\delta x(t) \equiv (\sum_{i=1}^N [x_i(t)-x_i(0)]^2/N)^{1/2}$, where the index $i$ labels individual stars). More precisely, Frankel et al. \cite{Frankel2020-vy} found that stars' angular momenta $J_\varphi$ have changed over the last $6$\,Gyr by
\begin{equation}
    \label{eqn:rmsJphi}
    \mathrm{rms} \,\, \delta J_\varphi \approx 619 \  \mathrm{kpc\, km\,s}^{-1}.
\end{equation}
Likewise, Frankel et al. \cite{Frankel2020-vy} found a total deviation in radial action close to the Sun over the last $6$\,Gyr of
\begin{equation}
    \label{eqn:rmsJR}
    \mathrm{rms} \,\, \delta J_R \approx 63 \, \mathrm{kpc \  km\,s}^{-1},
\end{equation}
though this quantity was found to increase weakly with Galactocentric radius.
Comparing \eqref{eqn:rmsJphi} with \eqref{eqn:rmsJR} we have that in the Milky Way over the last $6$ Gyr,
\begin{equation}
    \label{eqn:ratio}
    \frac{\mathrm{rms} \,\, \delta J_R}{\mathrm{rms} \,\, \delta J_\varphi} \approx 0.1.
\end{equation}

The quantities \eqref{eqn:rmsJphi} and \eqref{eqn:rmsJR} and especially their ratio \eqref{eqn:ratio} encode important information about the key drivers of dynamical evolution in our Galaxy. For instance, suppose stars underwent many impulsive kicks in velocity $\Delta \bm{v}$ (due to e.g., ISM or dark matter substructure) drawn from a random, homogeneous, isotropic distribution. This would drive migration on the order of $\mathrm{rms} \,\, \delta J_\varphi  \sim  R_\mathrm{g} \langle (\Delta \bm{v})^2\rangle^{1/2}$ and heating on the order of $\mathrm{rms}\,\, \delta J_R \sim {\langle (\Delta \bm{v})^2 \rangle}/{\kappa}$, where $\kappa$ is the local epicyclic frequency \citep{Binney1988-zy}.
Then
\begin{eqnarray}
   && \frac{\mathrm{rms} \, \delta J_R }{\mathrm{rms} \, \delta J_\varphi} \sim f \times \frac{\mathrm{rms} \, \delta J_\varphi}{J_\varphi}\nn 
   \\ 
&&\approx 2.5  \left( \frac{f}{7}\right) \left( \frac{\mathrm{rms} \, \delta J_\varphi }{619 \, \mathrm{kpc}\,\mathrm{km\,s}^{-1}}\right)  \left( \frac{J_\varphi}{1760 \, \mathrm{kpc}\,\mathrm{km\,s}^{-1}} \right)^{-1},\nn
\\
 \,
    \label{eqn:impulsive_ratio}
\end{eqnarray}
where $f$ is the product of various order-unity factors. Though we do not show any details here, we have confirmed the scaling $\mathrm{rms} \, \delta J_R \,/\, \mathrm{rms} \, \delta J_\varphi\propto \mathrm{rms} \, \delta J_\varphi$ using local shearing sheet simulations of stars interacting with isotropic, white noise Gaussian random fields of potential fluctuations with various power spectra. These simulations suggest $f\sim 7$, which would certainly produce too much heating-per-unit-migration compared to the measurement \eqref{eqn:ratio}. Thus, to the extent that such fluctuations do exist in real galaxies, the ratio \eqref{eqn:ratio} immediately places a strong constraint on their efficiency, since the large fractional heating they produce would need to be heavily diluted by other, `colder' mechanisms.

However, currently, there does not exist a quantitative theory that links the measurements \eqref{eqn:rmsJphi}-\eqref{eqn:ratio} to the spatio-temporal power spectrum of gravitational potential fluctuations in the disk, and in turn to specific perturbations like spiral arms, molecular clouds, dark matter substructure, etc. In a separate work (Hamilton, Modak \& Tremaine, in prep.), we have built upon the formalism of Hamilton et al. \cite{hamilton2024galactokinetics} to develop such a radial transport theory, and verified it numerically for perturbations of various amplitudes, wavelengths, frequencies, etc.\ in generic (thin) galactic disks. For the remainder of this paper we focus only on the Milky Way, and in particular on the effects of transient spiral structure, since this is believed to play a major role in the efficient radial migration of disk stars \citep{sellwood2022spirals}.

The community has focused so much on transient spirals because of the classic paper by Sellwood \& Binney \cite{Sellwood2002-lv}, hereafter referred to as SB02. 
These authors emphasized two important facts concerning star-spiral interactions.
First, provided the spiral lifetime $\tau$ is longer than the typical orbital period, the radial action $J_R$ of stars at the corotation resonance is adiabatically invariant.  Thus, resonant scattering at corotation can drive radial migration
 without radial heating.
Second, SB02 discussed how this corotation-scattering process works in the nonlinear regime. Precisely, the spiral will trap corotating stars onto `horseshoe' orbits, provided $\tau \gtrsim t_\mathrm{lib}/2$, where $t_\mathrm{lib}$ is the horseshoe libration period. For an $m$-armed spiral potential perturbation $\delta \phi$ with dimensionless amplitude $\eta \equiv |\delta\phi/(V_0^2 / 2)|$ (here, $V_0$ is the circular speed), this is (SB02; \cite{hamilton2024kinetic}):
\begin{align}
    t_\mathrm{lib} & \equiv \frac{t_\mathrm{dyn}}{m \, \eta^{1/2}} \nn \\
    \label{eqn:t_lib_CR}
    & \approx 0.64 \,\mathrm{Gyr} \, \left(\frac{m}{2}\right)^{-1} \left(\frac{\eta}{0.03} \right)^{-1/2} \left(\frac{t_\mathrm{dyn}}{220\,\mathrm{Myr}}\right),
\end{align}
where $t_\mathrm{dyn}$ is the orbital period at corotation. In this nonlinear horseshoe regime, when the spiral decays the trapped stars are `dropped off' with a new guiding radius, differing on average from the  initial guiding radius by about one resonance half-width $R_\mathrm{h}$, where
\begin{align}
    R_\mathrm{h} &\equiv \eta^{1/2} R_\mathrm{CR} \nn \\
    \label{eqn:half-radius}
    & \approx 1.4 \, \mathrm{kpc}\,\,\left(\frac{\eta}{0.03} \right)^{1/2} \left(\frac{R_\mathrm{CR}}{8\, \mathrm{kpc}} \right)
\end{align}
and $R_\mathrm{CR}$ denotes the corotation radius.

The smallness of the observed ratio \eqref{eqn:ratio} suggests that the first piece of SB02's argument, namely resonant scattering at corotation, does play a major role in transport in our Galaxy. However, the second piece of SB02's argument, namely nonlinear horseshoe behavior, suffers from a possible pitfall. Efficient radial migration seems to have occurred across a broad range of Galactocentric radii \citep{Frankel2020-vy, lian2022quantifying}, meaning we require not just one localized `horseshoe' event, but multiple such events spread across several disk scale-lengths. In each event, the spiral must \textit{not only} live long enough to trap stars and produce horseshoe libration, it must \textit{also} decay rapidly enough to make room for the next spiral. The latter requirement is important because if multiple spiral patterns are present simultaneously, the clean, coherent horseshoe behavior described above can be destroyed by resonance overlaps, which can drive efficient radial migration \citep{Minchev2010-la,rovskar2012radial} but also significant radial heating. In fact, 
Daniel et al. \cite{Daniel2019} showed that resonance overlaps (and associated heating) can even occur for a single spiral, between its corotation and ultraharmonic resonances, should its amplitude be large enough. Thus, SB02's horseshoe mechanism is rather delicate, and it is not clear whether it can produce all the radial migration \eqref{eqn:rmsJphi} without excessive heating above the observed level \eqref{eqn:ratio}.

In this paper, we address this question using numerical simulations of test-particle disks subject to transient spiral arm perturbations. Our fiducial spirals are designed to be morphologically similar to those observed in the Milky Way today \citep{Eilers2020-na}, but we also explore the effect of modifying each spiral's strength, pitch angle, number of arms, lifetime, and radial envelope, and also including a bar perturbation. 
We demonstrate that there are three distinct dynamical regimes of star-spiral interaction --- impulsive, resonant and horseshoe --- and that Milky-Way-like spirals in SB02's horseshoe regime are typically unable to reproduce the observations \eqref{eqn:rmsJphi} and \eqref{eqn:ratio}.
The data is better reproduced if one assumes spirals in the past were more open, more concentrated around corotation, shorter-lived, or some combination of these.
While we offer heuristic explanations for some of the trends we report here, we defer a rigorous treatment to a future comprehensive theory paper.

The remainder of this paper is organized as follows. In \S\ref{sec:Numerical_Setup}, we describe the setup of our test-particle simulations of the Galactic disk. In \S\ref{sec:Single_Spiral}, we illustrate the three key dynamical regimes that dictate the radial transport behavior
by performing controlled simulations involving a single transient spiral. In \S\ref{sec:Many_Spirals}, we simulate ensembles of many randomly generated spirals with various morphologies over $6$\,Gyr, and compare our results directly with the Milky Way data. In \S\ref{sec:Discussion}, we discuss the constraints that our results place on the spectrum of stochastic fluctuations present in our Galaxy over the last $6$\,Gyr. We summarize in \S\ref{sec:Summary}.

\section{Methods}
\label{sec:Numerical_Setup}

We consider an idealized, two-dimensional model of the Milky Way's stellar disk, employing standard polar coordinates $(\varphi, R)$.
We assume this disk is embedded in a rigid dark matter halo such that the axisymmetric part of the potential is logarithmic:
\begin{equation}
    \widetilde{\phi}(R) = V_0^2 \ln  \frac{R}{R_0},
    \label{eqn:log_halo}
\end{equation}
with $V_0 = 220$\,km\,s$^{-1}$ and $R_0=1$\,kpc. This produces a flat rotation curve with circular speed $V_0$ at all $R$. We treat the stellar population whose transport we are interested in as a set of test particles, which is justified as long as the drivers of transport --- in our case, transient spirals --- are much more massive than the stars themselves \citep{Binney1988-zy}. For simplicity, we do not include a galactic bulge, gas, dark matter substructure, or an evolving background potential. While such complications are all potentially important for radial transport, we believe that they will all systematically \textit{increase} the `heating-per-unit-migration' \inlineratio, the smallness of which is the primary concern of this paper. Because of this, if a proposed radial transport mechanism (e.g., horseshoe transport) cannot survive the relatively quiescent and idealized setup we consider here, it has little chance of doing so in the real Galaxy.

We subject our disk to perturbations by transient spirals of the form
\begin{equation}
    \label{eqn:log_spiral_transient}
    \delta \phi(\varphi, R, t) = A(t) B(R) \delta \phi^\mathrm{ls}(\varphi, R, t),
\end{equation}
where
\begin{align}
    \label{eqn:Phi_Spiral}
    &\delta \phi^\mathrm{ls}(\varphi, R, t)
    \nn \\ 
    & \equiv -\eta \frac{ V_0^2}{2} \cos\big[m\cot\alpha\ln \frac{R}{R_0} + m(\varphi-\pattern t-\varphi_\mathrm{p})\big],
\end{align}
is a logarithmic spiral with dimensionless strength $\eta$, $m$ arms, pitch angle $\alpha$, pattern speed $\pattern$ and azimuthal phase angle $\varphi_\mathrm{p}$.
The function 
\begin{equation}
    A(t) \equiv \me^{-(t-t_\mathrm{peak})^2/(2\tau^2)},
\end{equation}
sets the temporal envelope of the spiral to be a Gaussian peaked at some time $t_\mathrm{peak}$ and with characteristic lifetime $\tau$.
Meanwhile, the function
\begin{align}
    B(R) \equiv \me^{-(R-R_\mathrm{CR})^2/(2R_\beta^2)},
\,\,\,\,\,\,\,\,
    R_\beta \equiv \beta\frac{\sqrt{2} R_\mathrm{CR}}{m},
    \label{eqn:Radial_Envelope_Def}
\end{align}
sets the radial envelope of the spiral; it is peaked at the corotation radius $R_\mathrm{CR}\equiv V_0/\pattern$, and has standard deviation $R_\beta$ equal to $\beta$ times the distance, $\sqrt{2} R_\mathrm{CR}/m$, between corotation and the inner/outer Lindblad resonance (ILR/OLR). For instance, the choice $\beta = 1$ ($0.5$) corresponds to a spiral whose amplitude at Lindblad resonance is about $60\%$ ($13\%$) of that at corotation, meaning the \textit{power} in such spirals at Lindblad resonance is around $36\%$ ($1.7\%$) of that at corotation. The limit $\beta \to \infty$ corresponds to a radially-uniform envelope.


Eilers et al. \cite{Eilers2020-na} have fit an $m=2$ logarithmic spiral perturbation of the form \eqref{eqn:Phi_Spiral} to present-day Milky Way kinematics. They found a best-fit pitch angle of $\alpha = 12^\circ$ and a very extended radial envelope, which we idealize as $\beta \to \infty$.
Plugging Solar-neighborhood-like numbers into their equation (17), we can infer a dimensionless perturbation strength at the Solar radius of a few percent; for our fiducial value we take $\eta = 0.03$. We will use these parameters to guide our choices throughout this paper, but we explore the effect of varying all of them over a sensible range.

In addition to the transient spirals described above, in some of our simulations we add a simple rotating bar perturbation following Dehnen \cite{Dehnen2000}. The bar has radius $3.2$\,kpc, grows in strength smoothly over the first $2$\,Gyr of the simulation up to a final dimensionless strength $A_\mathrm{f} = 0.048$, and has a fixed pattern speed of $\Omega_\mathrm{b} = 35.2\,\mathrm{km\,s}^{-1}\,\mathrm{kpc}^{-1}$ (which corresponds to a corotation radius of $6.25\,\mathrm{kpc}$). These choices are roughly in line with Milky Way estimates based on GAIA DR3 \citep{Chiba2022-qt, Dillamore2023}; we have checked that our results are not sensitive to reasonable variations in the bar parameters.

For the initial conditions of the stars, we always choose random angles $\theta_\varphi \in [0, 2\pi)$ and $\theta_R \in [0, 2\pi)$, and random actions according to the Schwarzchild DF $f_0(\mathbf{J}) \propto \exp(-J_\varphi/\langle J_\varphi\rangle -J_R/\langle J_R\rangle)$, with $\langle J_\varphi\rangle = 4R_0V_0$ and $\langle J_R \rangle = 0.01R_0 V_0$. The latter choice corresponds to an initial radial velocity dispersion at radius $R$ of $\sigma \approx 9\,\mathrm{km\, s}^{-1}\,\times (R/8\,\mathrm{kpc})^{-1/2}$. In each simulation, we employ $N=10^4$ particles. We checked that quadrupling $\langle J_R \rangle$, using larger $N$, etc., did not materially affect our results.

\section{Results}

\subsection{A single transient spiral}
\label{sec:Single_Spiral}

In this section we perform controlled simulations involving a single spiral. We fix the spiral parameters $m=2$ and pitch angle $\alpha=12^\circ$, and we choose the pattern speed $\pattern$ such that corotation is at $R_\mathrm{CR}=8R_0 = 8$\,kpc. Below we report on such simulations with a variety of strengths $\eta$, envelope widths $\beta$, and lifetimes $\tau$.

\subsubsection{Three dynamical regimes}

\begin{figure*}
    \centering
    \includegraphics[width=0.8\textwidth]{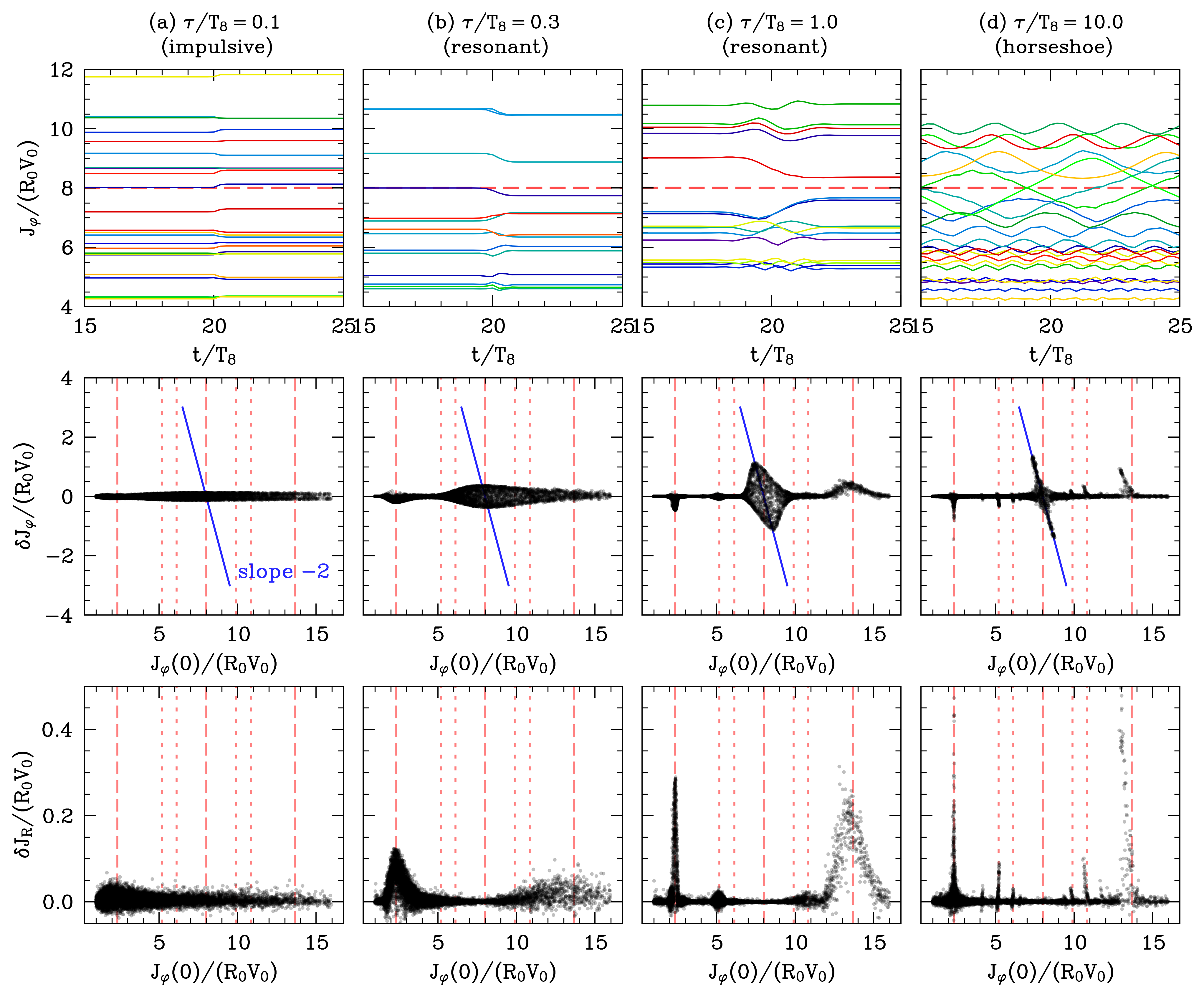}
    \caption{The response of the disk to a single $m=2$ spiral with pitch angle $\alpha=12^\circ$, corotation at $R_\mathrm{CR}=8.0R_0$, dimensionless strength $\eta=0.01$ and uniform radial envelope ($\beta\to\infty$). The four columns (a)-(d) correspond to spirals with different lifetimes $\tau$ as indicated in their labels. In the top panels, the red dashed horizontal line corresponds to corotation. In the other panels, the red dashed vertical lines correspond to corotation and Lindblad resonances, while dotted vertical red lines correspond to the $n= 2, 3$ UHR locations (equation \eqref{eqn:UHR_definition}). In each column, the top panel shows a random selection of $J_\varphi(t)$ trajectories from the simulation, and the middle (bottom) panels show the change in angular momentum $\delta J_\varphi$ (radial action $\delta J_R$) for each star at the end of the simulation as a function of its initial angular momentum $J_\varphi(0)$. A slope of $-2$ in the middle panels corresponds to stars `swapping places' across corotation, a classic sign of SB02's horseshoe mechanism (see column (d)). Note that for these spirals, $t_\mathrm{lib}/T_8 = 5.0$.}
    \label{fig:illustration1}
\end{figure*}

First, we set $\eta = 0.01$ and $\beta\to \infty$, and run four simulations corresponding to $\tau/T_8 = $ (a) $0.1$, (b) $0.3$, (c) $1.0$, (d) $10.0$, 
where $T_8 \equiv 2\pi(8R_0)/V_0 \approx 220$\,Myr is the circular period at $8$\,kpc (note that the libration period \eqref{eqn:t_lib_CR} in each case is $t_\mathrm{lib}/T_8 = 5.0$). We simulate for $40T_8$ and choose each spiral to peak in amplitude half way through the simulation, $t_\mathrm{peak}=20T_8$. We truncate the initial $J_\varphi$ distribution at $J_\varphi^\mathrm{min}=1R_0V_0$ and $J_\varphi^\mathrm{max}=16R_0V_0$, which is a wide enough range to include both of the spiral's Lindblad resonances.

In the top panels of Figure \ref{fig:illustration1} we show, for each of  these simulations, a random selection of the resulting $J_\varphi(t)$ timeseries around the peak time, $15 \leq t/T_8\leq 25$. In the middle (bottom) panels we show the corresponding change in angular momentum, $\delta J_\varphi$ (change in radial action, $\delta J_R$) of each star by the end of the simulation as a function of its initial angular momentum $J_\varphi(0)$. The dashed red horizontal line in the top panels corresponds to the corotation resonance location of a circular orbit, $J_\varphi = R_\mathrm{CR}V_0$. In the other panels, red vertical dashed lines denote the corotation resonance as well as the  Lindblad resonances, $J_\varphi = R_\mathrm{ILR/OLR}V_0$, while red vertical dotted lines correspond to ultraharmonic resonances (UHR), i.e., $J_\varphi = R_\mathrm{UHR}V_0$ where
\begin{equation}
    R_\mathrm{UHR} = \bigg( 1\pm \frac{\sqrt{2}}{mn} \bigg) R_\mathrm{CR},
    \label{eqn:UHR_definition}
\end{equation}
with $ n =  2, 3$. 

From the behavior exhibited in Figure \ref{fig:illustration1} we can identify three key dynamical regimes which we will refer to throughout this paper:
\begin{itemize}
    \item \underline{Impulsive regime}: \begin{equation}
        \tau \lesssim t_\mathrm{res}, \,\,\,\,\,\,\,\, \mathrm{where} \,\,\,\,\,\,\,\,\, t_\mathrm{res} \equiv \frac{t_\mathrm{dyn}}{2m}.
        \label{eqn:impulsive_def}
    \end{equation} 
    This is the case for almost all stars in the left column of Figure \ref{fig:illustration1}. In this regime, the spiral's lifetime is so short that stars with orbital periods $\sim t_\mathrm{dyn}$ cannot know that they are in a wave potential, and so resonances play no role. Each star receives a small impulsive kick whose magnitude is approximately proportional to $\tau$. Also, the torque each star feels depends approximately sinusoidally on its initial orbital phase, meaning the $\delta J_\varphi$ distribution at a fixed $J_\varphi(0)$ is nearly symmetric around zero.
    \item \underline{Resonant regime}: \begin{equation}
         t_\mathrm{res} \lesssim \tau \lesssim \frac{t_\mathrm{lib}}{2},
         \label{eqn:resonant_def}
         \end{equation}
         where $t_\mathrm{lib}$ is the libration time \eqref{eqn:t_lib_CR} evaluated at peak spiral amplitude. In Figure \ref{fig:illustration1}, $t_\mathrm{lib}/T_8=5.0$, meaning the resonant regime applies to columns (b) and (c). In this regime, stars that are not close to the spiral's corotation or Lindblad resonances feel a force that approximately averages to zero by the time the spiral decays. By contrast, the resonant stars feel a non-zero time-integrated force. The main effect is angular momentum transport at corotation and heating at the Lindblad resonances. However, the spiral still does not live long enough to produce horseshoe behavior.
    \item \underline{Horseshoe regime}: \begin{equation}
        \tau \gtrsim \frac{t_\mathrm{lib}}{2}.
                \label{eqn:horseshoe_def}
        \end{equation}
        This is the regime underlying SB02's mechanism; spirals that live at least this long give rise to horseshoe transport. A horseshoe orbit's $J_\varphi$ value crosses corotation and then crosses back again each libration period, as in column (d) of Figure \ref{fig:illustration1} (the slope of $-2$ in the middle panel corresponds to stars `swapping places' across corotation, and is a classic sign of horseshoe behavior --- see SB02). In the bottom two panels of this column we also observe transport at the UHRs --- see below.
\end{itemize}

\begin{figure*}
    \centering
    \includegraphics[width=0.8\textwidth]{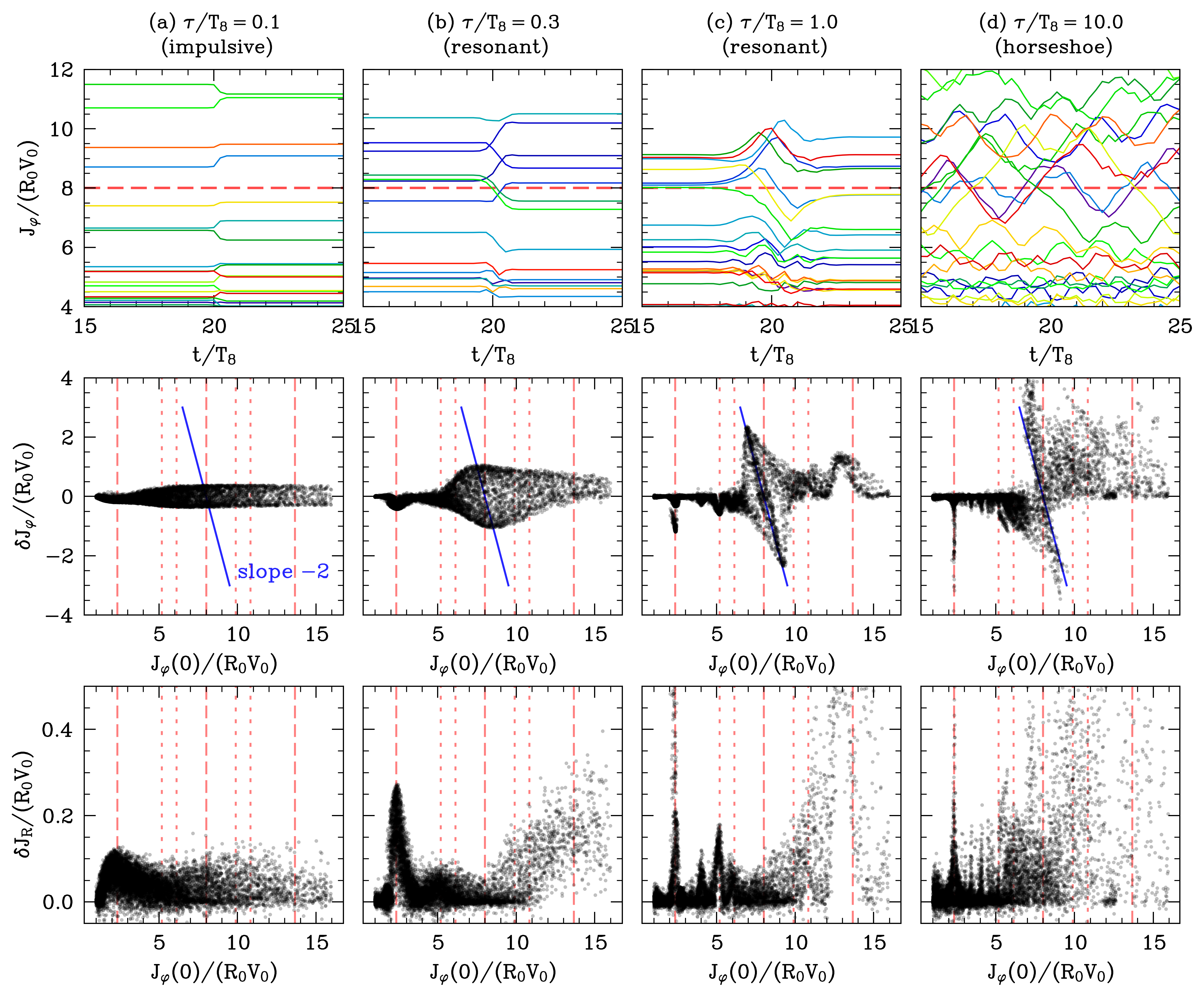}
    \caption{As in Figure \ref{fig:illustration1} except for a stronger spiral, with $\eta=0.03$. In this case, the libration time becomes $t_\mathrm{lib} = 2.9T_8$.}
    \label{fig:illustration2}
\end{figure*}

The behavior in the impulsive and resonant regimes exhibited in Figure \ref{fig:illustration1} (columns (a)-(c)) is rather generic. However, the behavior in the horseshoe regime can be quite different from that shown in column (d) of Figure \ref{fig:illustration1} if the spiral amplitude is large enough. To illustrate this, we present Figure \ref{fig:illustration2}, in which we performed the same simulations as in Figure \ref{fig:illustration1} except with $\eta = 0.03$, corresponding to a libration time $t_\mathrm{lib}/T_8 = 2.9$. We see from Figure \ref{fig:illustration2}d that now there is much more dramatic --- and much less sharply defined --- transport in the  $(J_\varphi(0), \delta J_\varphi)$ and $(J_\varphi(0), \delta J_R)$ planes in the horseshoe regime compared to the case with $\eta=0.01$ shown in Figure \ref{fig:illustration1}d. The main reason for this difference is that the corotation resonance is now wide enough (equation \eqref{eqn:half-radius}) to overlap with the UHR locations (vertical dotted lines in Figure \ref{fig:illustration2}). While isolated UHRs are not expected to drive significant transport on their own, resonance overlap between corotation and UHRs in the horseshoe regime leads to chaotic motion and enhanced transport, as first pointed out by Daniel et al. \cite{Daniel2019}. This phenomenon places a fundamental limit on the `coldness' of the nonlinear horseshoe mechanism at larger spiral amplitudes.

\subsubsection{Heating versus migration}

\begin{figure}
    \centering
    \includegraphics[width=0.49\textwidth]{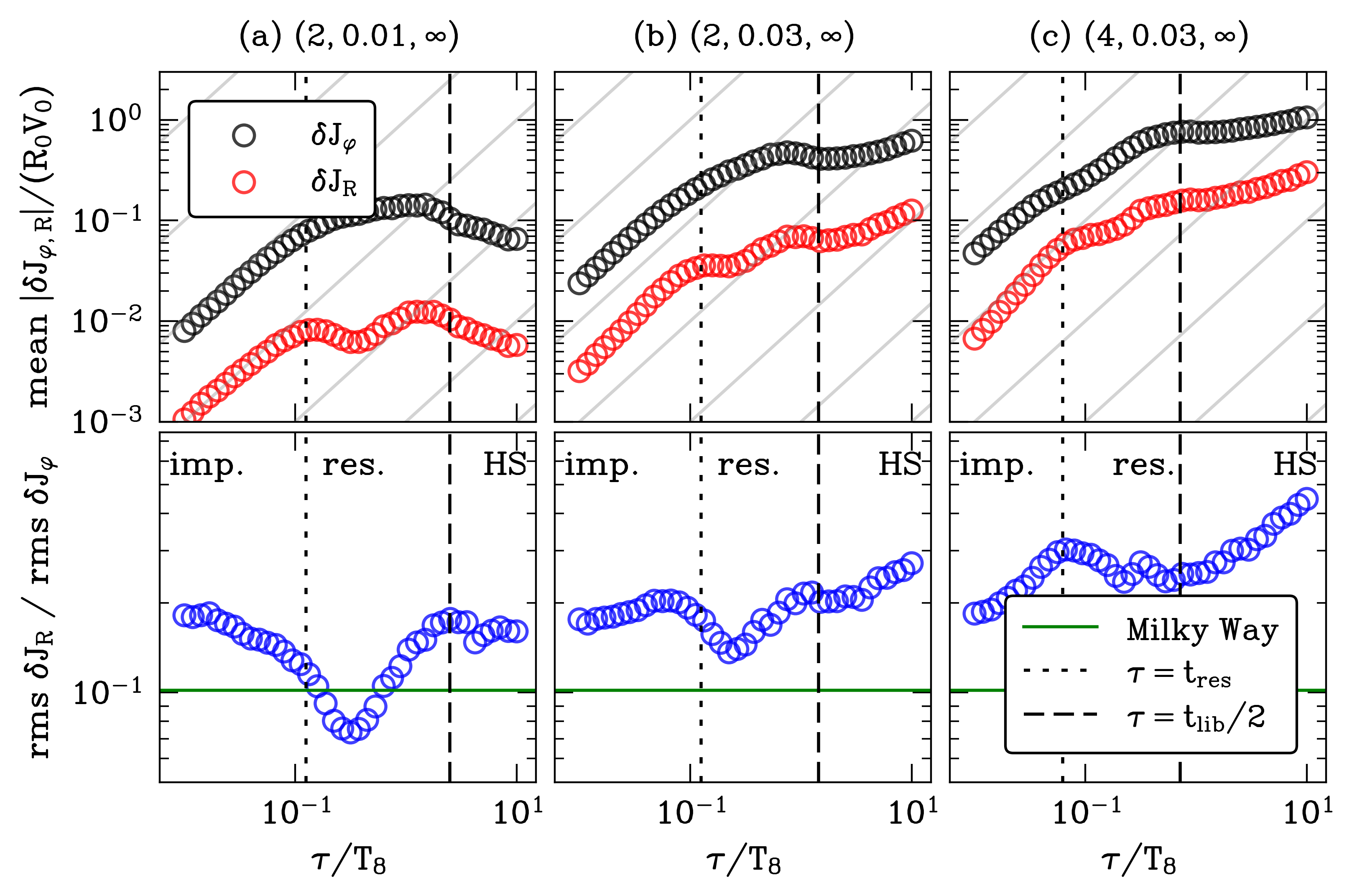}
    \caption{Mean absolute changes to $J_\varphi$ (black) and $J_R$ (red), and the ratio \inlineratio (blue),
    as a function of spiral lifetime $\tau$,
    for a disk perturbed by a single spiral with pitch angle $\alpha=12^\circ$ 
    and corotation radius at $R_\mathrm{CR} = 8R_0=8$ kpc. The label above each panel gives the remaining spiral parameters $(m, \eta, \beta)$. The vertical dotted and dashed lines correspond to $\tau = t_\mathrm{res}$ and $\tau=t_\mathrm{lib}/2$ respectively, and so separate the impulsive (`imp.', defined in equation \eqref{eqn:impulsive_def}), resonant (`res.', see equation \eqref{eqn:resonant_def}) and horseshoe (`HS', see equation \eqref{eqn:horseshoe_def}) regimes. The green horizontal line in the lower panels shows the measured value of \inlineratio in the Milky Way.}
    \label{fig:blackredblue}
\end{figure}

Next, we ran additional single-spiral simulations like the ones above, but for many different values of $\tau/T_8$, spanning all three dynamical regimes \eqref{eqn:impulsive_def}-\eqref{eqn:horseshoe_def}, and using three different choices of the parameters $(m, \eta, \beta)$, namely (a) $(2, 0.01, \infty)$ as in Figure \ref{fig:illustration1}; (b) $(2, 0.03, \infty)$ as in Figure \ref{fig:illustration2}; and (c) $(4, 0.03, \infty)$. Moreover, when applying our results to the Milky Way in the next section, we will not be interested in stars interior to about $4$\,kpc, since there the potential will be strongly distorted by the Galactic bar (and such stars are not part of the sample of \cite{Frankel2020-vy}). Hence, from now on we truncate the initial $J_\varphi$ distribution at $J_\varphi^\mathrm{min}=4R_0V_0$.

In Figure \ref{fig:blackredblue} we plot the resulting values of average absolute angular momentum change, mean $\vert  \delta J_\varphi \vert$ (black), and average absolute radial action change, mean $ \vert \delta J_R \vert$ (red), where the mean change in a quantity $x$ is deefined as mean\,$\vert \delta x(t) \vert \equiv \sum_{i=1}^N \vert x_i(t)-x_i(0)\vert /N$. In the bottom row of the figure, we also show the fractional heating-per-unit-migration \inlineratio (blue) as measured at the end of each simulation. The vertical dotted and dashed lines correspond to $\tau = t_\mathrm{res}$ and $\tau = t_\mathrm{lib}/2$ respectively, and hence separate the impulsive, resonant and horseshoe regimes \eqref{eqn:impulsive_def}-\eqref{eqn:horseshoe_def} described in the previous section. The diagonal lines in the upper panels show a linear scaling, mean $\vert \delta J_{\varphi, R} \vert \propto \tau$. Finally, the green line in the bottom panels shows the observed ratio \inlineratio in the Milky Way.

In every column (a)-(c) we see that, in the impulsive regime $\tau \lesssim t_\mathrm{res}$, both mean $\vert  \delta J_\varphi \vert$ and mean $\vert  \delta J_R \vert$ grow approximately linearly with $\tau$. This is as expected, since in the impulsive regime the force applied to each star is approximately constant, so the accumulated action change is proportional to the time for which this force is applied. More importantly, in all examples shown here, in the impulsive regime the value of \inlineratio is typically well above $0.1$. Even though these simulations only involve a single spiral, this rules out such perturbations as the primary drivers of heating and migration, since an accumulation of multiple such impulses cannot produce a smaller heating-to-migration ratio.

As $\tau$ is increased and we enter the resonant regime ($t_\mathrm{res}\lesssim \tau \lesssim t_\mathrm{lib}/2$), the ratio \inlineratio begins to decrease with $\tau$. We have found that the precise behavior in this resonant regime can be quite complicated depending on the exact location of the resonances, but often the ratio \inlineratio achieves its minimum here. Moreover, to the extent that only corotation and the ILR/OLR contribute significantly in this regime, the dynamics is roughly \textit{quasilinear} (see \S5.2 of \cite{hamilton2024galactokinetics}). In this regime a superposition of many such spirals should produce more migration and more heating, but roughly in the same ratio, i.e., without increasing
\inlineratio.

The same cannot be said for spirals in the horseshoe regime $\tau \gtrsim t_\mathrm{lib}/2$. As we enter this regime, we see from Figure \ref{fig:blackredblue}b and \ref{fig:blackredblue}c that the ratio \inlineratio increases with $\tau$, owing to the enhanced corotation-UHR overlap. This does not occur in Figure \ref{fig:blackredblue}a because there is no such overlap at smaller $\eta$ (recall Figure \ref{fig:illustration1}). In any case,
a superposition of many horseshoe-regime spirals would necessarily produce additional resonant overlaps and give rise to chaotic regions, leading to more heating-{per-unit-migration} than from a single spiral. This is a major disadvantage of spirals in the horseshoe regime: they cannot be superposed without increasing the ratio \inlineratio.

Furthermore, increasing the number of arms to $m = 4$ (column (c)) leads to a larger amount of overall radial migration, but also to more heating-per-unit-migration, because the ILR/OLR/UHR are closer to corotation, enhancing resonance overlap. Thus the single $m=4$ spiral is inconsistent with the data at every $\tau$.
\subsection{Many transient spirals}
\label{sec:Many_Spirals}

In this section, we run a suite of simulations in which the disk is randomly peppered with $N_\mathrm{sp}>1$ spirals of the form \eqref{eqn:log_spiral_transient}. All $N_\mathrm{sp}$ spirals within a given simulation have a fixed $(m, \eta, \beta, \alpha)$. Moreover, every time we generate such a spiral we choose a random value of its phase $\varphi_\mathrm{p}\in(0,2\pi]$, pattern speed $\pattern \in [V_0/(16R_0), V_0/(4R_0)]$ and peak time $t_\mathrm{peak} \in [0, 6\,\mathrm{Gyr}]$. For each set of fixed spiral parameters $(N_\mathrm{sp}, m, \eta, \beta, \alpha)$, we carry out simulations in which all spirals share the same lifetime $\tau$, as well as a simulation in which each spiral's lifetime is set to $\tau = 2\pi/\pattern$.

When $N_\mathrm{sp}$ is relatively small, the set of Monte-Carlo parameters $(\varphi_\mathrm{p}, \Omega_\mathrm{p}, t_\mathrm{peak})$ cannot be explored very thoroughly in any one simulation. Therefore, to ensure that our results are not just statistical flukes, for each set of fixed parameters we run an ensemble of $10$ simulations.

\subsubsection{Spirals with a uniform radial envelope ($\beta \to \infty$)}
\label{sec:beta_infty}

\begin{figure*}
    \centering
    \includegraphics[width=0.85\textwidth]{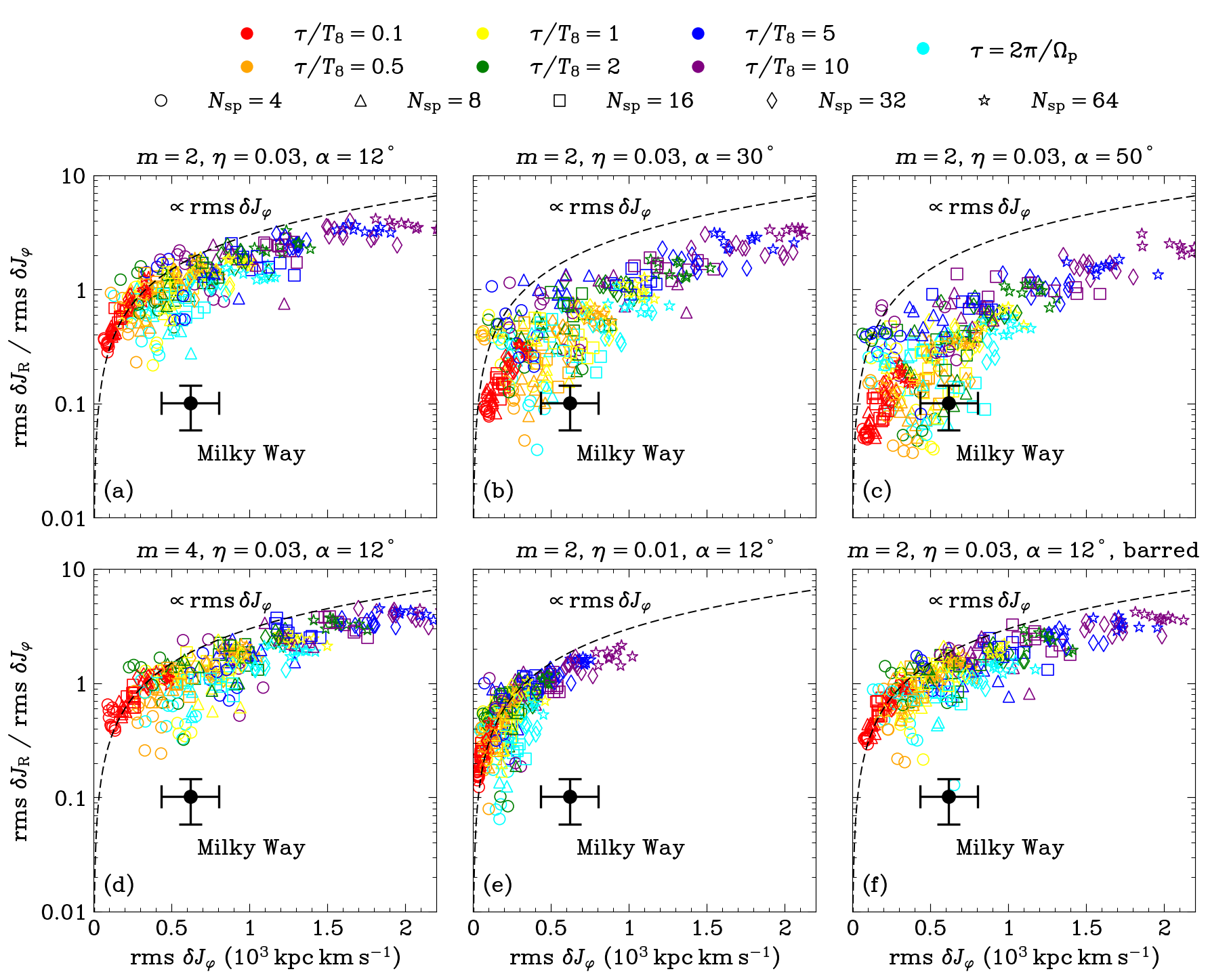}
    \caption{Radial migration and fractional heating-per-unit-migration after $6$\,Gyr of evolution for simulations with $N_\mathrm{sp}$ randomly generated spirals with a given $(m, \eta, \alpha)$ as indicated at the top of each panel, and $\beta \to \infty$ (i.e., no radial variation). The colored symbols represent $\tau/T_8 = 0.1$ (red), $0.5$ (orange), $1$ (yellow), $2$ (green), $5$ (blue), $10$ (purple); the cyan symbols indicate a simulation in which $\tau = 2\pi/\pattern$ for each individual spiral. Simulations with $N_\mathrm{sp} = 4$, $8$, $16$, $32$ and $64$ spirals are plotted as circles, triangles, squares, diamonds and stars respectively. The data point for the Milky Way is taken from the measurements \eqref{eqn:rmsJphi}-\eqref{eqn:rmsJR} with error bars of $\pm 30\%$ applied to each (see the Discussion). In panel (f), we include the perturbation due to the Galactic bar. In all panels we show with a black dashed line the scaling \eqref{eqn:impulsive_ratio}, setting $J_\varphi = 1760$ kpc km s$^{-1}$ and $f=5.2$.}
    \label{fig:m2}
\end{figure*}

First, we run simulations using (roughly) Milky-Way-like parameters: $m=2$, $\eta=0.03$, $\alpha=12^\circ$, and $\beta = \infty$. The remaining unfixed parameters are then the number of spirals $N_\mathrm{sp}$ and their characteristic lifetimes $\tau$, neither of which are known observationally.


In Figure \ref{fig:m2}a we show the radial migration and fractional heating-per-unit-migration for ensembles of simulations with $N_\mathrm{sp} = 4, 8, 16, 32, 64$ (circles, triangles, squares, diamonds, and stars respectively), and for $\tau/T_8 = 0.1$ (red), $0.5$ (orange), $1$ (yellow), $2$ (green), $5$ (blue), $10$ (purple). We show the results of the simulations in which $\tau = 2\pi/\pattern$ in cyan. We also plot the observed data point for the Milky Way using the measurements \eqref{eqn:rmsJphi}-\eqref{eqn:rmsJR} from \cite{Frankel2020-vy}, and assuming errors in both of these measurements of $\pm 30\%$ --- see \S\ref{sec:Discussion} for discussion.

None of the simulations reported in Figure \ref{fig:m2}a produce radial transport that is consistent with the Milky Way data, as might have been suspected from Figure \ref{fig:blackredblue}b. Remarkably, most of these simulations produce a ratio \inlineratio $\sim 1$, an order of magnitude larger than the observed ratio \eqref{eqn:ratio}. In fact, the black dashed line in this figure shows the scaling \eqref{eqn:impulsive_ratio} that we derived for isotropic scattering in \S\ref{sec:Introduction} (here setting $J_\varphi = 1760$ kpc\,km\,s$^{-1}$ and $f=5.2$). The fairly good fit suggests that when many spirals overlap one another, they behave effectively like uncorrelated, isotropic random noise.

It is possible that the pitch angles of spiral structure in the past were different from what is observed today. Therefore, next we ran the same simulations as in Figure \ref{fig:m2}a, except with $\alpha=12^\circ$ replaced by $\alpha = 30^\circ$ and $\alpha = 50^\circ$ respectively. Note that the radial wavelength at $R=8$ kpc of an $m=2$ logarithmic spiral with $\alpha=(12^\circ, 30^\circ, 50^\circ)$ is $\lambda = (2\pi R/m) \tan \alpha = (5.3, 15.5, 30)$\,kpc, so these new spirals are \textit{much} more open than those observed in the Milky Way today.
(Relatedly, \cite{yu2020statistical} analyzed $>4000$ spiral galaxies and found the vast majority had $\alpha\in(10^\circ, 30^\circ)$; almost none had  $\alpha \gtrsim 45^\circ$.). As Figures \ref{fig:m2}b and \ref{fig:m2}c show, a longer radial wavelength leads to a weaker radial forcing, hence a smaller ratio of heating-per-unit-migration. Because of this, some of the simulations --- typically those in the resonant regime, e.g., yellow and cyan symbols --- can be reconciled with the data, though 
these tend to be rather extreme cases of their respective ensembles. Spirals in the horseshoe regime (green, blue and purple) rarely match the Milky Way data even with these open pitch angles.

Next, some studies of Milky Way spirals have concluded that the structure is four-armed \citep{vallee2017guided}, rather than two-armed as assumed in \cite{Eilers2020-na}.
There may also have been a different number of arms in the past compared to today. To test the effect of varying $m$ we reran every simulation from Figure \ref{fig:m2}a, except now with $m=4$. The results of this exercise are shown in Figure \ref{fig:m2}d. As expected from the discussion of single-spiral transport in \S\ref{fig:blackredblue}, increasing $m$ enhances the radial migration rate
but also boosts the heating-per-unit-migration, well above the observed level.
Therefore, changes in arm number are not a promising way to design perturbations consistent with the data.

Similarly, measuring the amplitude of spirals is difficult, and there is no guarantee that spiral amplitudes have been the same over time. We know from \S\ref{sec:Single_Spiral} that increasing the value of $\eta$ will only tend to increase the amount of heating, so we rule out the possibility that significantly higher spiral amplitudes $\eta > 0.03$ could be reconciled with the data. To check the effect of smaller amplitudes, in Figure \ref{fig:m2}e we rerun  all simulations from Figure \ref{fig:m2}a except with  $\eta = 0.01$. This does help to reduce the fractional radial heating-per-unit-migration (as we might have expected from Figure \ref{fig:blackredblue}a). The main problem now is that these spirals do not produce enough overall migration. Notice also that the larger $N_\mathrm{sp}$ is, the larger \inlineratio becomes (i.e., star symbols tend to sit higher in this plot than diamonds, diamonds higher than squares, etc.) Hence, simply adding more weak spirals to make up for the lack of migration will inevitably lead to too much heating. We conclude that changes to spiral amplitude do not produce any straightforward reconciliation of theory and data.

Finally, the Milky Way is a barred galaxy and it has long been recognized that SB02's clean horseshoe mechanism might be modified in regions where the corotation resonance of a spiral transient overlaps with the bar's OLR \citep{Minchev2010-la}. In Figure \ref{fig:m2}d we show the results of rerunning all simulations from Figure \ref{fig:m2}a, now including the bar perturbation described in \S\ref{sec:Numerical_Setup}. We find that the bar perturbation does not have a particularly strong impact, presumably because it decays in amplitude rapidly beyond $\sim 4$\,kpc, and so is unable to affect the disk over multiple scale lengths. On the other hand, there is one simulation in this case that is consistent with the Milky Way data (denoted with a cyan circle). This should remind us that orbital transport is a stochastic process and it is possible, though not likely, that the Milky Way's dynamical history is just a statistical anomaly.


\subsubsection{Spirals with a finite radial envelope}

\begin{figure*}
    \centering
    \includegraphics[width=0.85\textwidth]{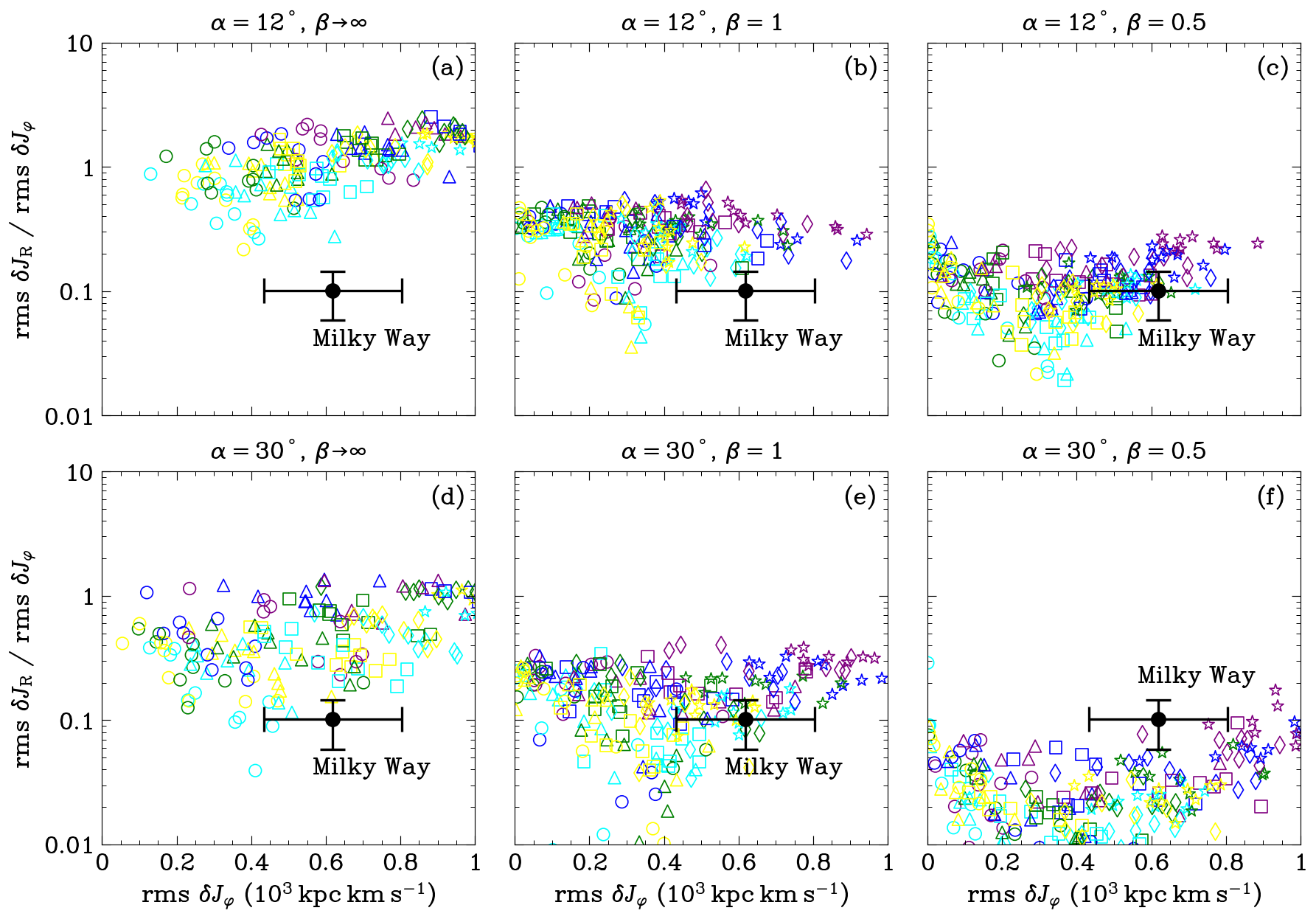}
    \caption{As in Figure \ref{fig:m2}a and \ref{fig:m2}b, with spiral parameters fixed to $(m, \eta) = (2, 0.03)$ and $\alpha = 12^\circ$ (panels (a)-(c)) and $30^\circ$ (panels (d)-(f)), but varying the radial envelope concentration parameter $\beta$. The choice $\beta = 1$ ($0.5$) corresponds to a Gaussian envelope in radius that is peaked at corotation and for which the Lindblad resonances are separated from the peak by one (two) standard deviation(s). The color and marker scheme is the same as in Figure \ref{fig:m2}, although for clarity we do not show any red/orange points, and the horizontal axis has a different scale.}
    \label{fig:m2_betas}
\end{figure*}

The final parameter we can sensibly vary is $\beta$, which sets the concentration of the radial envelope of the spiral around corotation (equation \eqref{eqn:Radial_Envelope_Def}). Most simulations of spiral-forming galactic disks find that radial envelopes are concentrated between inner and outer Lindblad resonances (ILR/OLR), with a peak somewhere in-between --- see for example Figure 10 of SB02, Figures 5 and 9 of \cite{sellwood2014transient}, and Figure 7 of \cite{vera2014effect}. To emulate these more concentrated profiles,we rerun the $\alpha = 12^\circ, 30^\circ$ simulations from Figure \ref{fig:m2}a and \ref{fig:m2}b, except with $\beta = 1$ and $\beta = 0.5$.

In panels (a) and (d) of Figure \ref{fig:m2_betas}, we simply reproduce the $\beta\to\infty$ results from
Figures \ref{fig:m2}a and \ref{fig:m2}b respectively (except that for clarity we no longer show red/orange points, and we have changed the horizontal scale so not all points are visible). The other panels show the effect of varying $\beta$. Unsurprisingly, decreasing $\beta$ (going from left to right in this figure) helps to reduce the ratio of heating-per-unit-migration, since it suppresses the UHR and Lindblad contributions (recall Figures \ref{fig:illustration1}-\ref{fig:illustration2}). Moreover, unlike in Figure \ref{fig:m2}, in these cases piling up more spirals (increasing $N_\mathrm{sp}$, i.e., going from circles to triangles to squares to diamonds to stars) does not lead to significantly more heating-per-unit-migration, since there are now far fewer resonance overlaps. In particular, the scaling \eqref{eqn:impulsive_ratio} is clearly not obeyed for $\beta \lesssim 1$.

We see from Figure \ref{fig:m2_betas}b that one can achieve near-agreement with the data by employing many Milky-Way-like ($\alpha=12^\circ$) 
spirals that are moderately concentrated around corotation ($\beta =1$, meaning the power in the spirals at Lindblad resonances is $36\%$ of that at corotation), with lifetimes in the resonant regime (cyan points). 
Spirals producing horseshoe motion (green-blue-purple) still drive too high a heating-to-migration ratio and so are still largely in tension with the data. Figure \ref{fig:m2_betas}c shows that we \textit{can}
match the data using $\alpha=12^\circ$ spirals in the horseshoe regime, but only if they are strongly concentrated around corotation ($\beta = 0.5$, so power at Lindblad resonances is just $1.7\%$ of that at corotation).
Figure \ref{fig:m2_betas}e shows that both resonant-regime and horseshoe-regime spirals are viable if we make them significantly more open than is observed today ($\alpha=30^\circ$) \textit{and} 
concentrate them weakly around corotation ($\beta =1$).
Finally, Figure \ref{fig:m2_betas}f shows that concentrating these more open spirals strongly around corotation ($\beta=0.5$) tends to overshoot the data, with \inlineratio falling below $0.1$ at almost all $\tau$. 

\section{Discussion}
\label{sec:Discussion}

Arguably the simplest question one can ask in non-equilibrium galactic disk dynamics is \textit{what drives orbital transport}? If that orbital transport consists of efficient radial migration without much radial heating, then the standard answer to this question is SB02's mechanism:
nonlinear horseshoe behavior at the corotation resonance of transient spiral structure.

We agree with SB02 that spirals' corotation resonances are crucial --- any mechanism that does not rely on these will tend to produce far too much heating per unit migration. However, our numerical experiments have shown that if the observed spiral morphology in the Galaxy today (especially its pitch angle $\alpha$) is representative of that over the past $6$\,Gyr, 
then the nonlinear horseshoe component of SB02's argument can only be reconciled with transport data (equations \eqref{eqn:rmsJphi}-\eqref{eqn:rmsJR}) if spirals are strongly suppressed away from corotation ($\beta \approx 0.5$, meaning the power at Lindblad resonances is just $\lesssim 1.7\%$ of that at corotation).
This seems to be in contradiction with the very uniform envelope of our Galaxy's spirals inferred by \cite{Eilers2020-na}, although measuring these is a highly uncertain task.
An alternative possibility is that the spirals in the past were much more open than they are observed to be today, perhaps with $\alpha \approx 30^\circ$.
In that case, either SB02 horseshoes or resonant (non-horseshoe) scattering
can be made consistent with the data, provided there is still some concentration of the spirals around corotation ($\beta \lesssim 1$).

Thus, we are not able to \textit{rule out} SB02's horseshoe mechanism as the main driver of radial transport in the Galaxy, but it has survived here only in a rather finely-tuned form. It is not clear whether this level of fine tuning is (or was) realistic, or what other observational signatures would allow us to confirm it.
At the same time, we should bear in mind that there is considerable dispersion in our results even for fixed spiral parameters, especially when including the effect of the Galactic bar (Figure \ref{fig:m2}f). The Milky Way is only one galaxy, and it is possible, though not likely, that its measured properties are a statistical anomaly drawn from a high-variance distribution.

However, the most important conclusion of this work is not that resonant scattering is preferred over horseshoe motion or vice versa. Instead, the crucial lesson is that \textit{we had to very work hard in order to design a set of perturbations that would produce transport consistent with the data}. In other words, reproducing \textit{both} the observed heating \textit{and} migration is a severely non-trivial requirement of Milky Way models (and, indeed, of the theory of spiral structure). It is also a good test of the veracity of `Milky Way analogue' galaxies identified in cosmological simulations --- if we require that such a simulated galaxy experiences the same amount of radial migration as the Milky Way, not just the same heating/thickening, we will greatly increase the probability that its life was truly `Milky-Way-like'. We recommend that simulators measure the ratio \eqref{eqn:ratio} whenever producing galaxies that are meant to resemble the Milky Way down to redshift zero. 


Since we are using the measurements  \eqref{eqn:rmsJphi}-\eqref{eqn:ratio} to make a rather serious claim about disk evolution, it is important to discuss whether there may simply be some significant error in those measurements. Yet, the radial migration efficiency \eqref{eqn:rmsJphi} has been agreed upon to within a few tens of percent in two separate models of Milky Way evolution \citep{Frankel2020-vy,lian2022quantifying}. Moreover, as noted in \cite{Frankel2020-vy}, it concurs nicely with a simple estimate based on the Solar neighborhood: from the metallicity gradient $\md [\mathrm{Fe/H}]/\md R \approx -0.06$ dex/kpc and the observed metallicity dispersion for old stars $\sigma_{[\mathrm{Fe/H}]} \approx 0.2$ dex, we can infer $\mathrm{rms} \,\, \delta J_\varphi \approx  \frac{\sigma_{[\mathrm{Fe/H}]} V_0}{\vert \md [\mathrm{Fe/H}]/\md R \vert} \approx 700 \, \mathrm{kpc \, km\,s}^{-1}$. Similarly, the estimate of radial heating \eqref{eqn:rmsJR} cannot be too far from the truth: assuming stars were born on circular orbits and follow an exponential distribution in $J_R$ today, we can use the present-day radial velocity dispersion $\sigma \approx 38$ km s$^{-1}$ and local epicyclic frequency $ \kappa \approx 37$ km\,s$^{-1}$kpc$^{-1}$ to estimate $\mathrm{rms} \,\, \delta J_R \approx \sqrt{2}\sigma^2/\kappa =55 \, \mathrm{kpc \, km\,s}^{-1}$. 
 In fact, in \cite{Frankel2020-vy}, the value of rms $\delta J_R$ was found to increase slightly with Galactocentric radius, with \eqref{eqn:rmsJR} being the appropriate value near the Sun.  Since there are many more disk stars inside the Solar radius than outside it, a true global measure of rms $\delta J_R$ might actually be smaller than \eqref{eqn:rmsJR}, meaning the ratio \eqref{eqn:ratio} might, if anything, be revised downwards.

Thus, we do not believe that the measurements \eqref{eqn:rmsJphi} and \eqref{eqn:rmsJR} can be wrong by more than a few tens of percent, and we have chosen the error bars in Figures \ref{fig:m2} and \ref{fig:m2_betas} to reflect this. These error bars would have to be drastically larger in order to alter our conclusions qualitatively.

\section{Summary}
\label{sec:Summary}

Recent observational advances have led to measurements of the radial transport of stars in the Milky Way's disk over the last $6$\,Gyr, revealing a large amount of radial migration (equation \eqref{eqn:rmsJphi}) and a remarkably small ratio of heating-per-unit-migration (equation \eqref{eqn:ratio}). We have tested the dependence of these radial transport properties of perturbations in the Galaxy, and in particular on the morphology, lifetime, amplitude, etc.\ of spiral structure. Our conclusions can be summarized as follows:
\begin{itemize}
    \item In general, it is very difficult to design perturbations that produce enough migration to match that observed in the Galaxy (equation \eqref{eqn:rmsJphi}) while not simultaneously producing too much radial heating, i.e., preserving the small ratio \eqref{eqn:ratio}.
    
    \item If the measured spiral structure in the Galaxy today is representative of that in the past, the classic SB02 horseshoe mechanism typically produces too much radial heating per unit radial migration to be consistent with observations. Bar-spiral resonance overlap does not help, and additional random scattering by disk or halo substructure is expected to make the problem worse (equation \eqref{eqn:impulsive_ratio}).
    
    \item 
    The data is better reproduced if one fine-tunes the spirals to be either much more open than is observed today, or much more concentrated around corotation, or some combination of these.
    Somewhat shorter-lived spirals (producing resonant, but not horseshoe, scattering) can also be reconciled with the data, and are perhaps preferable because they require less fine-tuning.
\end{itemize}
Most importantly, the measurements \eqref{eqn:rmsJphi}-\eqref{eqn:ratio} provide a very stringent test for models of radial transport in the Milky Way's disk over the second half of its life. They are also a major challenge to theories of spiral structure, and a good test of the veracity of `Milky Way analogue' galaxies extracted from large-volume cosmological simulations.

\begin{acknowledgments}
We thank Eve Ostriker for many helpful conversations, and Jerry Sellwood and Agris Kalnajs for comments on the manuscript. C.H. is supported by the John N. Bahcall Fellowship Fund and the Sivian Fund at the Institute for Advanced Study. S.M. acknowledges support from the National Science Foundation Graduate Research Fellowship under Grant No. DGE-2039656. The authors are pleased to acknowledge that the work reported on in this paper was substantially performed using the Princeton Research Computing resources at Princeton University which is consortium of groups led by the Princeton Institute for Computational Science and Engineering (PICSciE) and Office of Information Technology's Research Computing.
\end{acknowledgments}

%

\vspace{5mm}
\bibliographystyle{naturemag}
\bibliography{apssamp}

\end{document}